\documentclass[runningheads]{llncs}

\usepackage[T1]{fontenc}
\usepackage[utf8]{inputenc}   
\usepackage{lmodern}

\usepackage{amsmath,amssymb}
\usepackage{stmaryrd}         
\usepackage{newunicodechar}   

\newunicodechar{⇒}{\ensuremath{\Rightarrow}}    
\newunicodechar{→}{\ensuremath{\rightarrow}}    
\newunicodechar{⟹}{\ensuremath{\Longrightarrow}}
\newunicodechar{⟶}{\ensuremath{\longrightarrow}}
\newunicodechar{∀}{\ensuremath{\forall}}        
\newunicodechar{∈}{\ensuremath{\in}}            
\newunicodechar{∪}{\ensuremath{\cup}}           
\newunicodechar{¬}{\ensuremath{\neg}}           
\newunicodechar{∧}{\ensuremath{\wedge}}         
\newunicodechar{∨}{\ensuremath{\vee}}           
\newunicodechar{⦃}{\ensuremath{\lbrace}}        
\newunicodechar{⦄}{\ensuremath{\rbrace}}        
\newunicodechar{λ}{\ensuremath{\lambda}}        
\newunicodechar{⋀}{\ensuremath{\bigwedge}}      
\newunicodechar{⟦}{\ensuremath{\llbracket}}     
\newunicodechar{⟧}{\ensuremath{\rrbracket}}     

\DeclareUnicodeCharacter{2264}{\ensuremath{\le}}  
\DeclareUnicodeCharacter{2260}{\ensuremath{\neq}} 
\DeclareUnicodeCharacter{22C0}{\ensuremath{\bigwedge}} 
\DeclareUnicodeCharacter{21D2}{\ensuremath{\Longrightarrow}} 
\DeclareUnicodeCharacter{2200}{\ensuremath{\forall}} 
\DeclareUnicodeCharacter{2208}{\ensuremath{\in}} 
\DeclareUnicodeCharacter{03BB}{\ensuremath{\lambda}} 
\DeclareUnicodeCharacter{00AC}{\ensuremath{\neg}} 
\DeclareUnicodeCharacter{2983}{\ensuremath{\lbrace}} 
\DeclareUnicodeCharacter{2984}{\ensuremath{\rbrace}} 

\usepackage[T1]{fontenc}
\usepackage{float}
\usepackage{graphicx}

\usepackage{booktabs}
\usepackage{array}
\usepackage{makecell}
\usepackage{tabularx}
\usepackage{ragged2e}
\usepackage{url}

\usepackage{wrapfig}

\usepackage{url}

\usepackage[utf8]{inputenc} 
\usepackage[T1]{fontenc}    
\usepackage{hyperref}       
\usepackage{url}            
\usepackage{booktabs}       
\usepackage{amsfonts}       
\usepackage{nicefrac}       
\usepackage{microtype}      
\usepackage{xcolor}         

\usepackage[utf8]{inputenc}
\usepackage[T1]{fontenc}
\usepackage{microtype}
\usepackage{booktabs}
\usepackage{amsmath}
\usepackage{textcomp}
\usepackage{xcolor}
\usepackage{pgfpages}
\usepackage{pgfplots}
\usepackage{pgfplotstable}
\usepackage{tikz}
\usepackage{enumerate}
\usepackage[noend]{algpseudocode}
\usepackage{listings}
\usepackage{url}
\usepackage[multiple]{footmisc}
\usepackage{multirow}
\usepackage{arydshln}
\usepackage{xspace}
\usepackage{diagbox}
\usepackage{bm}
\usepackage{siunitx}
\usepackage{multirow}
\usepackage[breakable]{tcolorbox}

\usepackage[linesnumbered,ruled,vlined]{algorithm2e}
\usepackage{graphicx}
\usepackage{wrapfig}
\usepackage{balance}
\usepackage{caption}
\usepackage{subcaption}
\usepackage{array} 

\usepackage{amssymb,array,booktabs,tabularx}
\usepackage[T1]{fontenc}
\usepackage{bbm}
\usepackage{paralist}
\usepackage{xcolor,xspace}
\usepackage{minted}
\usepackage{booktabs}
\usepackage{array}
\usepackage{extarrows}
\usepackage[inline]{enumitem}
\usepackage[normalem]{ulem}
\usepackage{adjustbox}

\setminted{
	fontfamily=tt,
	fontsize=\footnotesize,
	numberblanklines=true,
	numbersep=12pt,
	numbersep=5pt,
	gobble=0,
	breaklines=true,
	frame=single,
	framerule=0.4pt,
	framesep=2mm,
	funcnamehighlighting=true,
	tabsize=4,
	obeytabs=false,
	mathescape=true,
	samepage=false, 
	showspaces=false,
	showtabs =false,
	escapeinside=@@,
	texcl=false,
}

\begin{document}
\title{Towards Real-World Industrial-Scale Verification: LLM-Driven Theorem Proving on seL4}
%
\titlerunning{LLM-Driven Theorem Proving for Real-World Verification}

\author{
Jianyu Zhang\inst{1} \and
Fuyuan Zhang\inst{1} \and
Jiayi Lu\inst{1} \and
Jilin Hu\inst{1} \and
Xiaoyi Yin\inst{1} \and
Long Zhang\inst{2} \and
Feng Yang\inst{2} \and
Yongwang Zhao\inst{1}\textsuperscript{*}
}

\authorrunning{J. Zhang et al.}

\institute{
Zhejiang University, Hangzhou, China
\and
National Key Laboratory of Science and Technology on Information System Security, China
}

%


\maketitle              

\begin{abstract}

Formal methods (FM) are reliable but costly to apply, often requiring years of expert effort in industrial-scale projects such as seL4, especially for theorem proving. Recent advances in large language models (LLMs) have made automated theorem proving increasingly feasible. However, most prior work focuses on mathematics-oriented benchmarks such as miniF2F, with limited evaluation on real-world verification projects. The few studies that consider industrial-scale verification mostly rely on closed-source models with hundreds of billions of parameters, which cannot be locally deployed and incur substantial usage costs. In this paper, we propose AutoReal, an LLM-driven theorem proving method for real-world industrial-scale systems with support for lightweight local deployment. We evaluate AutoReal on the seL4-Isabelle verification project as a representative and challenging case study. AutoReal incorporates two key improvements: (1) \textit{chain-of-thought (CoT)-based proof training}, which teaches the LLM the reasoning behind proof steps and enables step-wise explanations alongside proofs, and (2) \textit{context augmentation}, which leverages proof context from the project to enhance LLM-driven proving. Based on the AutoReal methodology, we fine-tune a base model to obtain AutoReal-Prover, a compact 7B-scale prover for industrial-scale theorem proving. AutoReal-Prover achieves a 51.67\% proof success rate on 660 theorems from seL4-designated Important Theories across all 10 seL4 proof categories, substantially outperforming prior attempts on seL4 (27.06\%). To evaluate generalization, we further apply AutoReal-Prover to three security-related projects from the Archive of Formal Proofs (AFP): \texttt{CRYSTALS-Kyber\_Security}, \texttt{RSAPSS}, and \texttt{Elliptic\_Curves\_Group\_Law}, covering all 451 theorems and achieving a proof success rate of 53.88\%. Overall, this work advances the application of LLM-driven theorem proving in real-world industrial-scale verification.


\end{abstract}
%
%
%

\section{Introduction}\label{sec:introduction}

Formal methods (FM) provide rigorous techniques for verifying critical systems, with theorem proving as a key approach \cite{bjorner201440,cao2024preface}.
Theorem proving establishes system correctness through machine-checked logical proofs and plays a central role in large-scale, high-assurance verification \cite{krichen2023survey,nawaz2019survey,hasan2015formal}. However, the human effort required to apply theorem proving remains prohibitively high, particularly for real-world industrial-scale projects. For example, the original seL4 verification reported about 20 person-years of proof effort~\cite{klein2009sel4}. Similarly, projects such as CompCert~\cite{leroy2016compcert} and CertiKOS~\cite{gu2016certikos} typically require multiple person-years of expert effort. These substantial costs pose a major barrier to the widespread adoption of formal methods.

Recent advances in large language models (LLMs) have demonstrated strong reasoning capabilities across a wide range of tasks~\cite{naveed2025comprehensive,shi2025continual,vaswani2017attention}, including model checking~\cite{wu2024llm,pirzada2024llm} and auto formalization~\cite{cao2024preface,weng2025autoformalization}. These advances motivate their application to automated theorem proving. Polu et al.\ explored the possibility of using LLMs to automatically complete theorem proving tasks~\cite{polu2020generative}. Following the release of evaluation benchmarks such as the mathematics-oriented miniF2F~\cite{zheng2021minif2f} dataset, substantial research efforts have focused on improving proof success rates on miniF2F, including LEGO-Prover~\cite{wang2023lego}, ProofAug~\cite{liu2025proofaug}, and Seed-Prover~\cite{chen2025seed}. These works have continually advanced state-of-the-art proof success rates on miniF2F. However, existing work has largely focused on specific benchmarks, with miniF2F serving as the primary evaluation benchmark~\cite{zheng2021minif2f}, which mainly consists of mathematical theorems and differs from verification tasks arising in industrial-scale systems. Extending these approaches to real-world industrial-scale verification projects remains challenging, as such projects involve complex system safety properties that typically require longer proof chains and substantially richer proof contexts~\cite{zhang2024selene,lin2024fvel}.

Few recent studies have begun to explore the use of LLMs for theorem proving in real-world verification projects \cite{zhang2024selene,lin2024fvel,qin2025can,bayazit2025case}. For the seL4 verification project, Selene~\cite{zhang2024selene} and FVELER~\cite{lin2024fvel} investigate the application of LLMs to discharging proofs in the seL4 verification project, and these works have demonstrated promising potential. Selene reports a best averaged proof success rate of 27.06\% with GPT-4 on 340 selected seL4 theorems, representing the previously best reported result on seL4~\cite{zhang2024selene}. However, applying LLM-driven theorem proving to real-world verification workflows still requires further improvements in proof success rates. Moreover, LLM-generated proofs may lack transparency for human understanding, as the reasoning behind individual proof steps is not made explicit~\cite{limperg2025tactic,lebese2021proof,tithi2025promise}. This raises additional concerns when deploying LLM-driven automation in safety-critical systems. In addition, prior works that consider industrial-scale verification often rely on closed-source models such as GPT-4~\cite{zhang2024selene,qin2025can,bayazit2025case,curaba2cryptoformaleval}, which operate at the scale of hundreds of billions of parameters, cannot be locally deployed, and incur substantial API costs when used for large-scale proof generation. In contrast, practical industrial verification typically requires lightweight local deployment due to operational, security, and cost constraints~\cite{paloniemi2025porting,pan2025cost,ye2024enabling}. To make LLM-driven theorem proving viable for real-world industrial-scale verification, it is therefore necessary to improve both proof success rates and provide explicit explanations of generated proofs. At the same time, practical verification environments call for models with small parameter scales that can be deployed locally under lightweight computational requirements.

To address the challenges, we propose AutoReal, an LLM-driven theorem proving method tailored for real-world industrial-scale verification workflows. AutoReal is designed to improve proof success rates while making the reasoning behind generated proofs explicit. To this end, AutoReal incorporates two improvements: CoT-based proof training and context augmentation. We instantiate AutoReal by fine-tuning a 7B-scale model to obtain AutoReal-Prover, a prover specialized for real-world verification projects that supports local deployment with lightweight computational requirements. We evaluate AutoReal-Prover on seL4, a representative and challenging industrial-scale verification project~\cite{heiser2020sel4_Australia}. We further evaluate it on security-related projects from the Archive of Formal Proofs (AFP)~\cite{mackenzie2021evaluation} to assess generalization. Our results show that AutoReal-Prover substantially improves proof success rates over prior work and provides step-aligned explanations with generated proofs. Our work advances the applicability of LLM-driven theorem proving to real-world industrial-scale verification.

In summary, the main contributions of this paper are as follows:

\begin{itemize}

\item \textbf{An LLM-driven method for real-world verification.}
We propose AutoReal, an LLM-driven theorem proving method for real-world industrial-scale verification. AutoReal introduces two key improvements: CoT-based proof training, which teaches the model the reasoning behind proof steps and enables explanation generation, and context augmentation, which supplies proof context from the project to support LLM-driven proving. Through these improvements, AutoReal contributes to extending LLM-driven theorem proving from benchmark-oriented datasets to real-world industrial-scale verification projects.

\item \textbf{A lightweight model for industrial-scale theorem proving.}
Based on the AutoReal methodology, we fine-tune a 7B-scale base model (Qwen2.5-Coder-7B~\cite{hui2024qwen2}) to obtain AutoReal-Prover, a model specialized for theorem proving in industrial-scale verification. Its training leverages an intermediate proof CoT dataset of about 200k step-level instances constructed in this study. In addition to improved proof success rates, AutoReal-Prover generates step-wise explanations alongside proofs to support human understanding and inspection. As a compact 7B-scale model, AutoReal-Prover enables cost-efficient local deployment, making it well suited for real-world industrial verification settings. AutoReal-Prover is the concrete outcome of our approach and has been released as an open-source model.\footnote{The model is available at \url{https://doi.org/10.5281/zenodo.18412012}.}

\item \textbf{Improved performance on real-world verification tasks.}
We evaluate AutoReal-Prover on real-world verification projects comprising diverse, unfiltered theorems across multiple proof categories. On seL4, it achieves a 51.67\% proof success rate on all 660 theorems from seL4-designated Important Theories~\cite{sel4_access_2026}, covering all 10 proof categories~\cite{noauthor_sel4_2026}, improving upon the previously reported best result of 27.06\%. We further demonstrate generalization on three security-related AFP projects~\cite{noauthor_archive_2026}, covering all 451 theorems and achieving a 53.88\% success rate. Overall, these results highlight the potential of LLM-driven theorem proving for real-world verification.

\end{itemize}

\paragraph{Outline.} The remainder of this paper is organized as follows.
\autoref{sec:background-motivation} introduces the verification workflow, illustrates where AutoReal fits in industrial verification, and presents a running example.
\autoref{sec:method} describes the design and implementation of AutoReal, including the method overview and its core methodological components.
\autoref{sec:experiment} presents a comprehensive experimental evaluation of AutoReal.
\autoref{sec:related} reviews related work.
Finally, \autoref{sec:conclusion} summarizes the contributions and future work of this study.


\section{Background: Verification Workflows and Use Cases}
\label{sec:background-motivation}

This section introduces the seL4 verification workflow and the application scenario of AutoReal. We first review how proofs are developed in seL4, and then describe where AutoReal fits into this workflow, followed by a concrete running example.

\subsection{seL4 Verification Workflow}\label{seL4_workflow_context}

\paragraph{Industrial-scale verification in seL4.}
seL4 is a high-assurance microkernel whose high-performance C implementation has been proven functionally correct in Isabelle/HOL, via a refinement-based argument from abstract specification down to C code~\cite{klein2009sel4,heiser2020sel4}.
The verified kernel comprises 8.7~KLOC of C code and about 600 lines of assembly, and is supported by a large proof base consisting of roughly 200{,}000 lines of Isabelle scripts, which has been maintained and extended alongside kernel evolution~\cite{klein2009sel4,heiser2020sel4}.
In the official seL4 verification repository, proofs are organized into ten categories that cover diverse verification tasks~\cite{noauthor_sel4_2026}, such as access control proof~\cite{sel4_access_2026} and assembly refinement proof~\cite{sel4_assembly_nodate}.

\paragraph{seL4 verification workflow and proof context.}
As illustrated on the left of Figure~\ref{fig0}, seL4 verification follows a layered refinement methodology:
it connects abstract and executable kernel specifications to the C implementation via refinement proofs~\cite{klein2009sel4,klein2010refinement}.
This refinement-based verification results in a large collection of theorems that must be proved in Isabelle/HOL~\cite{klein2010refinement}.
These theorems are not proved in isolation. To discharge a target theorem, the proof effort is typically decomposed into smaller obligations, requiring auxiliary lemmas and unfolding relevant definitions~\cite{klein2009sel4}.
Consequently, proofs rely on a substantial body of supporting material, primarily definitions and previously proved lemmas~\cite{klein2010refinement}.
As the development evolves, this body of supporting material steadily grows and continues to support subsequent theorems~\cite{klein2010refinement}.
We refer to this theorem-relevant auxiliary information as the \emph{proof context}.

\subsection{AutoReal: Application Scenario and Running Example}

\paragraph{Motivation: automating context-dependent interactive proofs.}
Although Isabelle provides powerful automation, proof development in industrial-scale projects such as seL4 remains largely human-driven~\cite{klein2009sel4,heiser2020sel4}.
For a given theorem, verifiers work under the available proof context, iteratively selecting proof strategies and composing proof steps until all subgoals are discharged.
This process is incremental and context-dependent. To prove a target theorem, verifiers often need to introduce and prove auxiliary lemmas and unfold relevant definitions, sometimes through multiple levels of decomposition~\cite{klein2010refinement}.
Once established, these auxiliary results are incorporated into the evolving proof context, supporting the target theorem and enabling reuse in subsequent proofs.

Constructing such proofs remains technically demanding and typically requires substantial effort from formal-methods experts~\cite{klein2009sel4}.
Moreover, as the kernel evolves, verification introduces new proof obligations and necessitates ongoing proof maintenance, further increasing the overall human cost~\cite{heiser2020sel4}.

\begin{figure}
\centering
\includegraphics[width=0.9\textwidth]{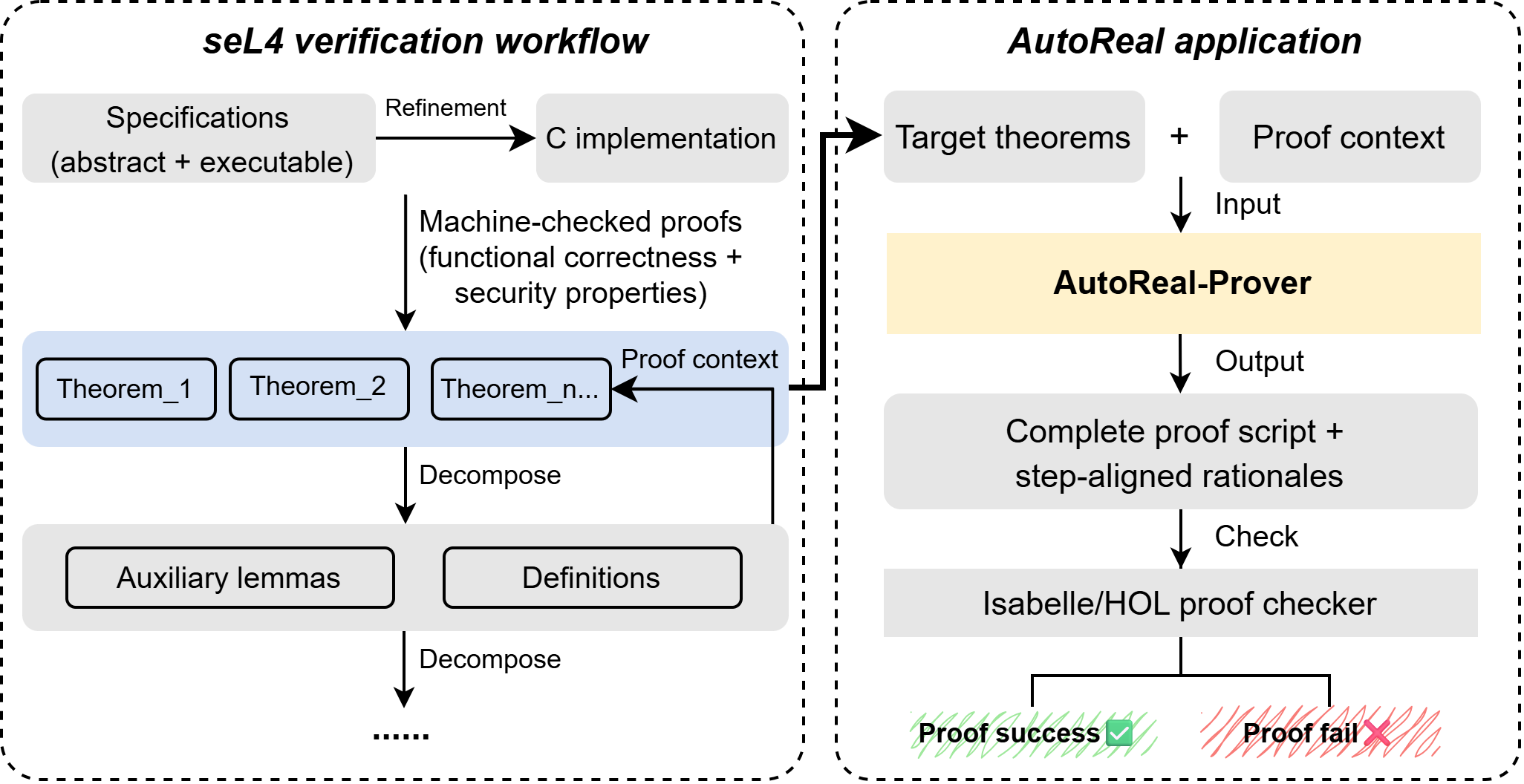}
\caption{The application scenario of AutoReal in the seL4 verification workflow.}
\label{fig0}
\end{figure}

\paragraph{Where AutoReal fits in the workflow.}
Figure~\ref{fig0} shows the application scenario of AutoReal.
Given a target theorem together with its proof context, AutoReal-Prover automatically synthesizes a complete Isabelle proof script.
To further support human inspection and repair, it also outputs step-aligned natural-language rationales alongside proof steps.

The resulting proof script is validated by the Isabelle/HOL proof checker.
If accepted, the proof can be reused in the development, reducing manual effort in both initial proof construction and proof maintenance as the kernel evolves.

\begin{figure}
\centering
\includegraphics[width=0.85\textwidth]{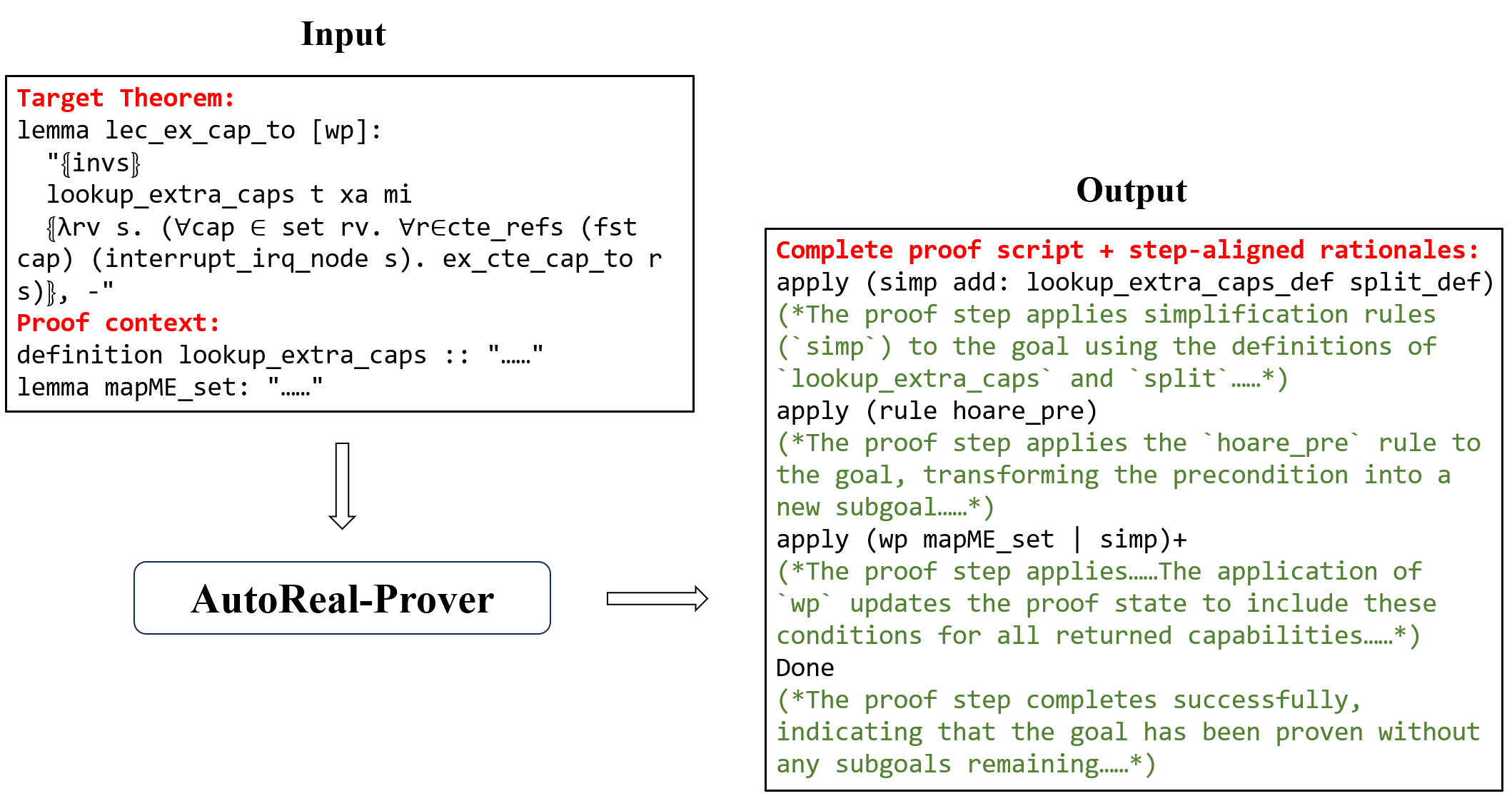}
\caption{A running example for AutoReal.}
\label{fig0-1}
\end{figure}

\paragraph{Running example.}\label{Running_example} Figure~\ref{fig0-1} shows a use case of AutoReal. The example is taken from the invariant-abstract proof category of seL4~\cite{sel4_abstract_nodate}. The target theorem \texttt{lec\_ex\_cap\_to[wp]} proves that all capabilities returned by \texttt{lookup\_extra\_caps} are supported by a corresponding capability-table entry in the abstract state. The input to AutoReal-Prover is the target theorem together with its proof context, here the definition of \texttt{lookup\_extra\_caps} and the traversal lemma \texttt{mapME\_set}. Using this context, AutoReal-Prover produces a proof script that unfolds the computation and propagates the required postcondition through the traversal. The resulting script is accepted by Isabelle, illustrating that AutoReal-Prover can effectively leverage the proof context to construct a correct proof.

Alongside the proof script, AutoReal-Prover also emits step-aligned natural-language rationales to explain the reasoning behind each proof step; a complete example of the generated rationales is shown in Fig.~\ref{fig:successful_proof_rationales} in the Appendix.

\section{Methodology}\label{sec:method}

In this section, we present the design and implementation of AutoReal. We first provide an overview of the proposed method, and then describe its main methodological components.

\subsection{Overview}
\begin{figure}
  \centering
  \includegraphics[width=1\textwidth]{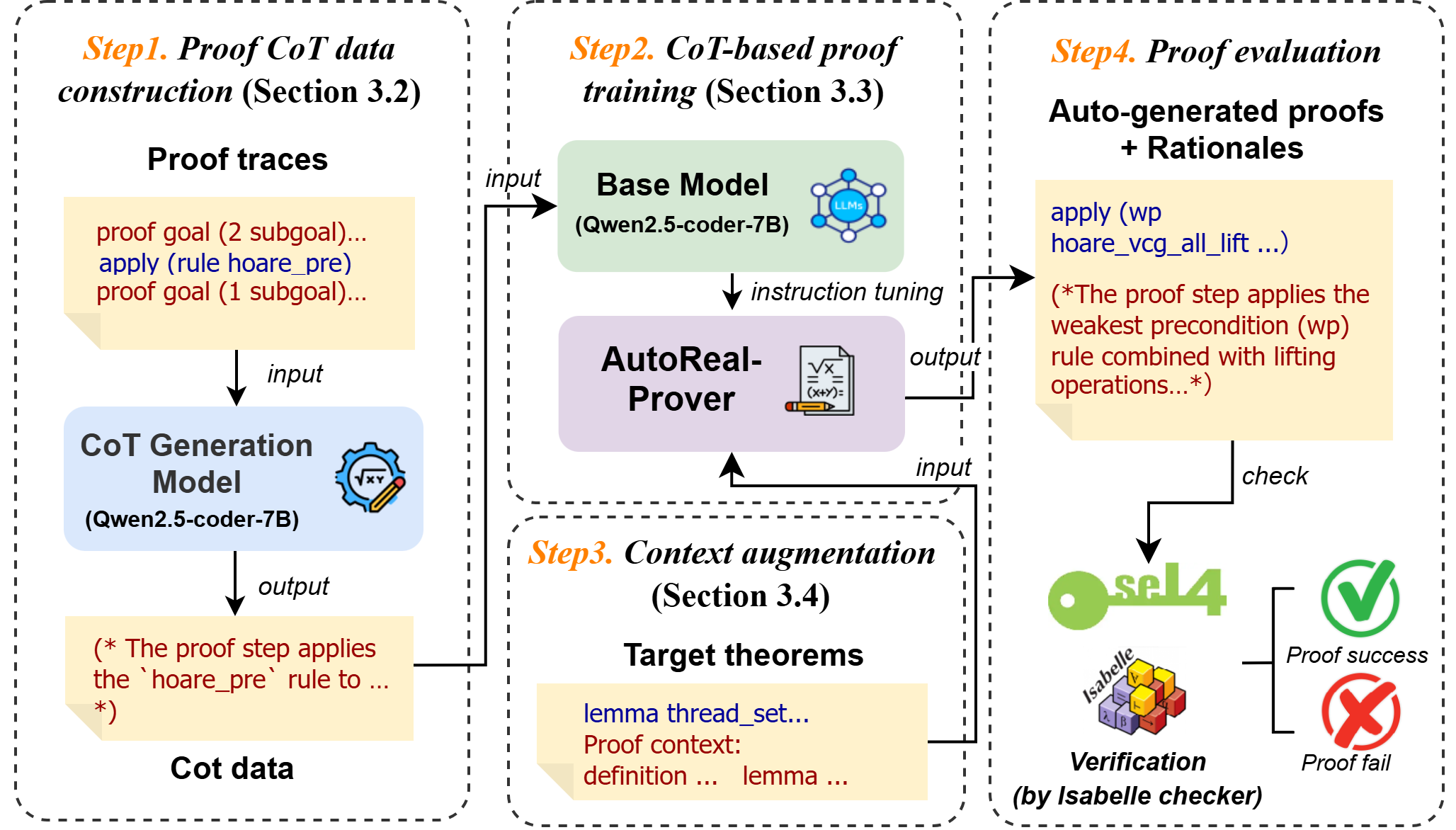}
  \caption{Method overview of AutoReal}
  \label{fig1}
\end{figure}

The method overview of AutoReal is shown in Fig.~\ref{fig1}. 
The method consists of four stages:
\begin{inparaenum}[\textit{Step}~1.]
\item \textit{Proof CoT data construction.}
We construct a proof CoT dataset from existing seL4 proofs, pairing individual proof steps with explanations of their underlying reasoning to make step-level proof reasoning explicit (Section~\ref{CoT_data_construction}).
\item \textit{CoT-based proof training.}
Using the constructed proof CoT data, we fine-tune a base model to obtain AutoReal-Prover, which generates complete proof scripts together with step-aligned explanations, enabling state-aware proof construction without repeated interaction with Isabelle (Section~\ref{proof_training}).
\item \textit{Context augmentation.}
During proof generation, AutoReal-Prover is supplied with theorem-relevant proof context, including auxiliary lemmas and definitions, allowing it to reason under the same assumptions as human verifiers (Section~\ref{Context_augmentation}).
\item \textit{Proof evaluation.}
AutoReal-Prover is applied to automatically generate proof scripts and rationales for new theorems using a predefined prompt template (Appendix~\ref{appendix:prompt_proof_generation}); the generated scripts are checked by the Isabelle proof checker, and a proof is considered successful if it is accepted without errors, solves all subgoals, and contains no placeholder commands such as \texttt{sorry} or \texttt{oops}~\cite{nipkow2002isabelle}.
\end{inparaenum}


\subsection{Proof CoT data construction}\label{CoT_data_construction}

\paragraph{Why step-level CoT for industrial-scale proofs.}
Step-level proof CoT data refers to associating each proof command with an explanation of its effect on the proof state.
This formulation aligns with the interactive theorem proving process, which proceeds by transforming a proof state through a sequence of proof steps, each operating on the current subgoals and assumptions.
In industrial-scale developments such as seL4, this process is highly context-sensitive and relies on recurring, project-specific proof patterns.
Consequently, understanding a proof requires not only knowing which commands are applied, but also how each step changes the proof state and enables subsequent reasoning.

Step-level proof CoT data makes this fine-grained reasoning explicit by associating each proof command with an explanation of its effect on the proof state. This step-level representation is essential for modeling how proofs are constructed in large verification projects and for producing explanations that are meaningful to human verifiers. To our knowledge, no public dataset provides step-level CoT rationales for industrial-scale Isabelle proofs; we therefore construct such a dataset from seL4.

\begin{figure}
\begin{center}
  \includegraphics[width=0.8\textwidth]{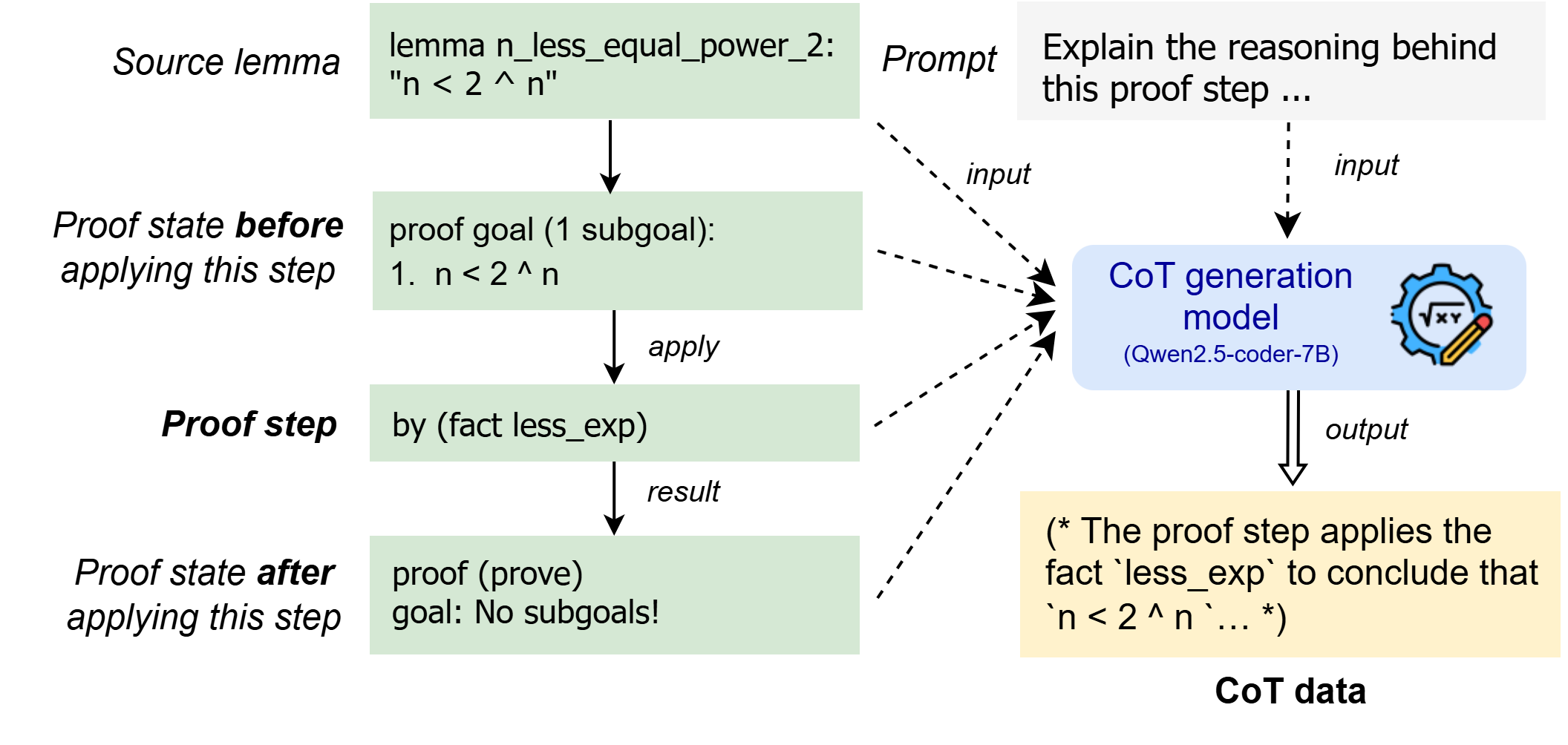}
  \caption{Step-level CoT data construction from Isabelle proof traces.}
  \label{fig3}
\end{center}
\end{figure}

\paragraph{Source proof traces and data split protocol.}
We construct step-level CoT data from proof traces extracted by FVELER~\cite{lin2024fvel}, which records seL4 proof developments as sequences of individual proof steps together with their corresponding Isabelle proof states.  Each trace associates a proof step with the proof state immediately before and after its execution, precisely characterizing the effect of the step on the current proof obligations. We analyze these proof traces using a CoT generation model to produce step-level proof CoT data, as shown in Figure~\ref{fig3}.

To prevent information leakage during evaluation, we adopt a strict hold-out protocol: proof steps associated with the theorems used for evaluation are excluded from CoT data construction and reserved for testing. As a result, the CoT training data contains no proof steps from the evaluation targets.

\paragraph{Structured input for CoT generation.}
For each proof step, we construct a structured input for CoT generation that combines four components: (1) the source lemma, including its identifier and statement; (2) the proof state before applying the step, including the current subgoals; (3) the proof step itself, namely the Isabelle command to be explained; and (4) the proof state after applying the step. Figure~\ref{fig3} illustrates this organization. Together, these components provide an explicit witness of how each proof step transforms the proof state.

Each structured input is paired with an instruction prompt that asks the model to explain the reasoning behind the proof step and its role in the larger proof, producing a state-aware explanation of why the step is applicable and how it contributes to proof progress. The full prompt is provided in Appendix~\ref{appendix:prompt_cot_construction}.

\paragraph{CoT generation model requirements.}
The CoT generation model should possess sufficient mathematical reasoning capability and familiarity with formal languages and interactive proof artifacts, including Isabelle proof commands and proof states. As recent LLM training corpora increasingly cover code, mathematics, and formal reasoning, these requirements are met by a growing number of open-source models. In our experiments, we use Qwen2.5-Coder-7B~\cite{hui2024qwen2} as the CoT generation model, due to its strong mathematical reasoning capability and compact parameter scale.

\paragraph{Generation procedure and resulting CoT instances.}
We feed each structured input into the CoT generation model and collect the resulting step-aligned rationales, yielding approximately 200k CoT instances from seL4 proofs. Each instance pairs a single proof step with a natural-language explanation describing (1) the proof methods, lemmas, and definitions applied in the step and their purpose; (2) the resulting change to the proof state, such as simplifying or discharging subgoals; and (3) the contribution of the step to the overall progress of the proof. A complete example of a CoT instance is provided in Appendix~\ref{appendix:cot_instance}.

\subsection{CoT-based proof training}\label{proof_training}

\paragraph{Training objective.}
The goal of CoT-based proof training is to enable AutoReal-Prover to construct complete and correct proofs in a single inference. In addition to generating correct proof scripts, the model is trained to produce natural-language rationales aligned with individual proof steps.

In interactive theorem proving, each proof step is selected based on its applicability to the current proof state. Training a model solely on proof scripts provides no explicit supervision on this state-dependent reasoning and therefore fails to capture how proof states evolve across steps. A straightforward alternative is to generate proofs incrementally while repeatedly querying Isabelle for updated proof states. However, such tightly coupled interaction incurs substantial computational overhead, particularly in large-scale developments such as seL4.

Our approach instead trains the model, via supervised learning on proof CoT data, to internalize proof-state evolution during proof construction. By learning from step-aligned explanations that justify why a step is applicable and how it advances the proof, AutoReal-Prover can reason about proof states without requiring repeated interaction with the Isabelle proof checker.

\paragraph{Proof-state transitions and CoT supervision.}
This training design is grounded in the structure of Isabelle/HOL proof developments. In Isabelle, the proof state maintains a list of remaining subgoals for the target theorem and guides the selection of subsequent proof steps~\cite{wenzel2007isabelle}. Internally, each subgoal is represented as a theorem in sequent form, typically written as
$A_1 \Longrightarrow \cdots \Longrightarrow A_n \Longrightarrow C$,
where $A_1,\ldots,A_n$ are the local assumptions/premises accumulated for that subgoal and $C$ is its current conclusion; $n = 0$ indicates a premise-free subgoal, and a proof is fully discharged when no subgoals remain. Proof methods transform the current proof state by decomposing, rewriting, or eliminating subgoals until all are resolved~\cite{wenzel2014isabelle}.

Rather than conditioning the model on the full proof state at every step, CoT-based proof training can be viewed as providing explicit reasoning traces as intermediate supervision~\cite{wei2022chain}. From a modeling perspective, chain-of-thought reasoning introduces an intermediate trace $Z = (z_1,\ldots,z_T)$ between the model input $x$ and the final output $y$, where each $z_t$ explains the intent of the corresponding proof step $a_t$ and how it is expected to affect the proof obligations~\cite{wei2022chain,nye2021show}. For a proof command sequence $A = (a_1,\ldots,a_T)$, applying $a_t$ transforms the proof state $G_{t-1}$ into $G_t$, yielding the goal-state evolution
$G_0 \xrightarrow{a_1} G_1 \xrightarrow{a_2} \cdots \xrightarrow{a_T} G_T$.
Accordingly, the structured trace used in training can be written as $(x, G_0) \Rightarrow (z_1, a_1) \Rightarrow \cdots \Rightarrow (z_T, a_T).$

During instruction tuning, the model input consists of the target lemma statement together with the relevant proof context, denoted by $x$, which induces the initial proof state $G_0$. The model is trained to generate the entire proof as a single sequence interleaving proof steps with chain-of-thought explanations, and generation terminates when the produced script fully discharges the goals, with the final explanation reflecting this completion.

\paragraph{Training data representation.}
Guided by the above modeling perspective, Figure~\ref{fig4} illustrates the structure of a training instance used for CoT-based proof training. The training input consists of the target lemma together with its associated proof context, while the training output is a single structured sequence that alternates between Isabelle proof commands and step-aligned reasoning explanations derived from the step-level proof CoT data constructed in Section~\ref{CoT_data_construction}.

\begin{figure}[H]
\centering
\textsc{Training Input (Lemma Statement + Proof Context)}\\[-0.6em]
\rule{0.92\linewidth}{0.2pt}
\begin{minted}[
  frame=none,
  fontsize=\fontsize{8}{9}\selectfont,
  breaklines,
  tabsize=2
]{isabelle}
lemma set_ep_bitmapQ_no_L1_orphans[wp]: "..."
proof context: "..."
\end{minted}

\vspace{0.1em}
\textsc{Training Output (Proof Step + Rationale)}\\[-0.6em]
\rule{0.92\linewidth}{0.2pt}
\begin{minted}[
  frame=none,
  fontsize=\fontsize{8}{9}\selectfont,
  breaklines,
  tabsize=2
]{isabelle}
whole proof:
apply (unfold setEndpoint_def)
(* The proof step unfolds the definition of `setEndpoint`, simplifying the goal without changing its logical meaning ... *)
apply (rule setObject_ep_pre)
(* The proof step applies the `setObject_ep_pre` rule, strengthening the precondition with the required invariant ... *)
apply (simp add: bitmapQ_defs setObject_def split_def)
(* The proof step applies simplification rules to reduce the complexity of the goal ... *)
apply (wp hoare_Ball_helper hoare_vcg_all_lift updateObject_default_inv | simp add: bitmapQ_def)+
(* The proof step applies WP-based reasoning to discharge the remaining proof obligations ... *)
done
(* The proof completes successfully with no remaining subgoals ... *)
\end{minted}

\vspace{-0.6em}
\caption{Training instance structure for CoT-based proof training.}
\label{fig4}
\end{figure}

For each proof command, the accompanying explanation describes how the step transforms the current proof state and advances the proof. This design explicitly trains the model to generate proofs together with aligned natural-language rationales and encourages it to connect successive proof steps by internally reasoning about proof-state transitions. As a result, step-aligned explanations can stand in for explicit proof-state queries, enabling proof generation in a single inference without repeated interaction with Isabelle.

\paragraph{Training implementation.}
We realize this training objective via instruction tuning of a base model to obtain AutoReal-Prover. In our implementation, we fine-tune Qwen2.5-Coder-7B~\cite{hui2024qwen2} using the training instances constructed as described in Figure~\ref{fig4}. This Qwen2.5-Coder-7B is chosen for its strong capability in code semantic understanding and reasoning~\cite{hui2024qwen2}, while its relatively compact 7B parameter scale facilitates cost-efficient local deployment in real-world verification settings. Each training instance prompts the model to generate the entire proof as a single sequence that interleaves proof commands with their corresponding CoT explanations. Detailed training configurations are provided in Appendix~\ref{appendix:training_details}.

\subsection{Context augmentation}\label{Context_augmentation}

\paragraph{Why context augmentation for industrial proofs.}
As discussed in Section~\ref{seL4_workflow_context}, proofs in industrial Isabelle developments rely on a substantial proof context accumulated throughout the verification process.
In large systems such as seL4, this context is rich and distributed across many theory files, which makes it difficult to complete proofs without explicitly identifying and supplying the relevant dependencies for a target theorem~\cite{klein2009sel4,klein2010refinement}.
This motivates the need for explicit context augmentation when applying automated theorem proving techniques in industrial-scale verification.

Prior work typically addresses this dependence by providing all theory files transitively required by the target theorem~\cite{zhang2024selene}.
Although automated, this approach becomes inefficient in large developments, where dependency chains may span dozens of layers~\cite{lin2024fvel}.
As a result, the prover is burdened with substantial token overhead and large amounts of information that are not directly relevant to the current proof obligation, which can obscure the essential structure of the problem and hinder effective reasoning.

\paragraph{Targeted context augmentation.}\label{context_analyze}
AutoReal adopts a targeted context augmentation strategy that mirrors how proofs are developed in large-scale verification projects.
In developments such as seL4, proofs are carried out under an evolving proof context that accumulates definitions and lemmas.
For a given theorem, verifiers select and reuse the available proof context to discharge the goal.

Following this workflow, AutoReal expects theorem-relevant proof context to be explicitly provided by the verifier together with the target theorem.
Instead of supplying entire theory files, a compact set of definitions and lemmas that are necessary for understanding the theorem statement and completing the proof is supplied as proof context.

Figure~\ref{fig0-1} illustrates how AutoReal-Prover exploits proof context during proof construction.
In this example, the proof context includes the definition of \texttt{lookup\_extra\_caps} (via \texttt{lookup\_extra\_caps\_def}), allowing AutoReal-Prover to unfold the function and expose the monadic, list-processing structure needed for weakest-precondition reasoning.
The lemma \texttt{mapME\_set} then connects the \texttt{mapME}-based traversal to a set-of-results view, enabling the WP step \texttt{(wp mapME\_set \textbar\ simp)+} to propagate the desired postcondition uniformly to all capabilities returned in the result list.
Without this supplied proof context, the proof would typically require manually recovering the key unfolding and appropriate traversal lemmas before automation can carry the postcondition through the computation, which is difficult in practice for both LLMs and human verifiers.

\paragraph{Context augmentation in our evaluation.}
In our evaluation, the proof context supplied to AutoReal-Prover is kept consistent with the definitions and lemmas used in the corresponding human-written proofs~\cite{noauthor_sel4_2026,noauthor_archive_2026}.
By aligning the provided context with the original proofs, we ensure that the auxiliary information is correct and sufficient for the target theorem, and that the evaluation focuses on automated proof construction rather than on discovering relevant context.
This setup isolates the effectiveness of context-aware proof generation.
It allows us to assess whether AutoReal can successfully exploit theorem-relevant context to synthesize complete proofs and aligned explanations, under the same assumptions and dependencies as those used in large-scale industrial verification practice.

\section{Experimental Evaluation}\label{sec:experiment}

In this section, we present an experimental evaluation of AutoReal-Prover. We first describe the evaluation objectives and setup, and then report quantitative results on the seL4 verification project~\cite{noauthor_sel4_2026} and on security-related projects from the AFP~\cite{noauthor_archive_2026}. We further provide qualitative analyses of the generated proofs and their natural-language rationales, followed by an ablation study assessing the impact of the introduced improvements.

Our experiments are designed to address the following research questions:

\begin{itemize}
\item \textbf{RQ1:} How does AutoReal-Prover perform across Important Theories in different seL4 proof categories?
\item \textbf{RQ2:} Does the proving capability of AutoReal-Prover generalize to other real-world verification projects?
\item \textbf{RQ3:} Are the natural-language rationales generated by AutoReal-Prover accurate and useful for human verifiers?
\item \textbf{RQ4:} Do the proofs generated by AutoReal-Prover differ from the original human-written proofs?
\item \textbf{RQ5:} Are the introduced improvements responsible for the observed gains in proof success rates? (ablation study)
\end{itemize}

\subsection{Evaluation Setup}

\subsubsection{Evaluation Projects and Theorems}

We evaluate AutoReal-Prover on the seL4 verification project and on additional real-world verification projects drawn from the AFP. Evaluation is conducted at the theorem level and focuses on complete theories used in real-world verification efforts.

\paragraph{seL4 and Important Theories.}
Our primary evaluation is performed on the seL4-Isabelle verification project. Since seL4 is actively maintained, we conduct the evaluation on the version released on March~1,~2024~\cite{noauthor_sel4_2026}, together with the corresponding Isabelle version specified by the project.

According to the official organization of the seL4 repository~\cite{noauthor_sel4_2026}, proofs are grouped into ten proof categories, each corresponding to a class of verification tasks, as summarized in Table~\ref{tab1}. For each category, the corresponding README file in the seL4 repository designates a set of \emph{Important Theories} that serve as the key proof developments for that category~\cite{sel4_abstract_nodate,sel4_assembly_nodate,sel4_access_2026}. Following the project guidance, for each proof category we select the first theory listed under \emph{Important Theories} in the README and evaluate AutoReal-Prover on all theorems contained in that theory, without cherry-picking. In the \texttt{capDL-api} category~\cite{sel4_capdl_2026}, the designated theories \texttt{API\_DP.thy} and \texttt{Kernel\_DP.thy} contain no theorems; we therefore evaluate \texttt{KHeap\_DP.thy}, which has the largest number of theorems in this category. None of the evaluated theorems are included in the CoT training data. A detailed description of the proof categories and the theories used in our evaluation is provided in the Appendix~\ref{appendix:sel4-proof-categories}.

\begin{table}
\caption{seL4 proof categories and the Important Theories used in evaluation.}
\label{tab1}
\centering
\small
\setlength{\tabcolsep}{4pt} 
\begin{tabularx}{\linewidth}{@{} >{\RaggedRight\arraybackslash}p{3.0cm}  >{\RaggedRight\arraybackslash}X  >{\raggedleft\arraybackslash}p{3.7cm} @{}}
\toprule
\textbf{Proof category} & \textbf{Description} & \textbf{Important theory} \\
\midrule
\texttt{access-control}     & Access Control Proof                                & \nolinkurl{Syscall_AC.thy} \\
\texttt{asmrefine}          & Assembly Refinement Proof                            & \nolinkurl{SEL4GraphRefine.thy} \\
\texttt{bisim}              & Bisimilarity of seL4 with a static Separation Kernel & \nolinkurl{Separation.thy} \\
\texttt{capDL-api}          & CapDL API Proofs                                     & \nolinkurl{KHeap_DP.thy} \\
\texttt{crefine}            & C Refinement Proof                                   & \nolinkurl{Refine_C.thy} \\
\texttt{drefine}            & CapDL Refinement Proof                               & \nolinkurl{Refine_D.thy} \\
\texttt{infoflow}           & Confidentiality Proof                                & \nolinkurl{Noninterference.thy} \\
\texttt{invariant-abstract} & Abstract Spec Invariant Proof                        & \nolinkurl{Syscall_AI.thy} \\
\texttt{refine}             & Design Spec Refinement Proof                         & \nolinkurl{Refine.thy} \\
\texttt{sep-capDL}          & CapDL Separation Logic Proof                         & \nolinkurl{AbstractSeparation_SD.thy} \\
\bottomrule
\end{tabularx}
\end{table}

\paragraph{AFP projects.}
To assess the generalization of AutoReal-Prover beyond seL4, we additionally evaluate it on security-related verification projects drawn from the AFP~\cite{noauthor_archive_2026}, a large collection of verification developments mechanically checked in Isabelle~\cite{mackenzie2021evaluation}. We use the stable AFP release from January~2026 together with the Isabelle2025-1 proof checker (December~2025)~\cite{noauthor_archive_2026,noauthor_isabelle-2025-1_2025}.

We consider three security-related projects: \texttt{CRYSTALS\mbox{-}Kyber\allowbreak\_Security}~\cite{CRYSTALS-Kyber-Security-AFP} (64 theorems in 9 theories), \texttt{RSAPSS}~\cite{RSAPSS-AFP} (278 theorems in 12 theories), and \texttt{Elliptic\_Curves\_Group\_Law}~\cite{Elliptic-Curves-Group-Law-AFP} (109 theorems in 3 theories). For each project, AutoReal-Prover attempts to prove all theorems under the same evaluation protocol as used for seL4. Additional details on these projects and the reasons for their selection are provided in Appendix~\ref{appendix:afp-projects}.

\subsubsection{Metric: Proof success rate}
The proof success rate measures the proportion of theorems that AutoReal-Prover can successfully prove within a theory or an entire project. To evaluate this metric, we replace the original human-written proofs with proofs generated by AutoReal-Prover in the corresponding theories and check them using the Isabelle proof checker. A theorem is considered successfully proved if at least one of up to 256 proof attempts is accepted by Isabelle, which means that the generated proof script discharges all subgoals without errors and contains no placeholder commands such as \texttt{sorry} or \texttt{oops}. This metric reflects the proving capability of AutoReal-Prover on real-world verification tasks.

\subsection{Experimental Results and Discussion}

\subsubsection{Results on seL4 (RQ1)}

Table~\ref{tab:sel4-results} summarizes the proof success rates of AutoReal-Prover on seL4.
Overall, AutoReal-Prover successfully proves 341 out of 660 theorems from the Important Theories, yielding an overall proof success rate of 51.67\%.
This result shows that AutoReal-Prover can automatically discharge a substantial portion of real-world seL4 verification obligations.
As shown in the table, proof success rates vary across different proof categories, reflecting differences in the complexity of the verification tasks.

\begin{table}
\caption{Proof success rates of AutoReal-Prover on seL4 across proof categories.}
\label{tab:sel4-results}
\centering
\small
\begin{tabularx}{\linewidth}{@{} l @{\hspace{1em}} >{\RaggedRight\arraybackslash}X >{\raggedleft\arraybackslash}p{1.8cm} >{\raggedleft\arraybackslash}p{3cm} @{}}
\toprule
\textbf{Proof category} & \textbf{Theory} & \textbf{Solved/Total} & \makecell[r]{\textbf{Proof success}\\\textbf{rate}} \\
\midrule
\texttt{access-control}     & \texttt{Syscall\_AC.thy}            & 47/84   & 55.95\% \\
\texttt{asmrefine}          & \texttt{SEL4GraphRefine.thy}        & 1/1     & 100\%   \\
\texttt{bisim}              & \texttt{Separation.thy}             & 5/9     & 55.56\% \\
\texttt{crefine}            & \texttt{Refine\_C.thy}              & 13/46   & 28.26\% \\
\texttt{drefine}            & \texttt{Refine\_D.thy}              & 2/3     & 66.67\% \\
\texttt{capDL-api}          & \texttt{KHeap\_DP.thy}              & 89/123  & 72.36\% \\
\texttt{infoflow}           & \texttt{Noninterference.thy}        & 101/233 & 43.35\% \\
\texttt{invariant-abstract} & \texttt{Syscall\_AI.thy}            & 57/105  & 54.29\% \\
\texttt{refine}             & \texttt{Refine.thy}                 & 26/55   & 47.27\% \\
\texttt{sep-capDL}          & \texttt{AbstractSeparation\_SD.thy} & 0/1     & 0\%     \\
\midrule
\textbf{Overall}            & --                                   & \textbf{341/660} & \textbf{51.67\%} \\
\bottomrule
\end{tabularx}
\end{table}

Direct comparison with prior work on seL4 is difficult, as the only existing LLM-driven theorem proving efforts on seL4 adopt different theorem selection strategies and evaluation protocols.
For example, Selene evaluates on 340 sampled seL4 theorems drawn from multiple sessions using predefined filtering and stratified sampling criteria, and reports a best averaged proof success rate of 27.06\%~\cite{zhang2024selene}. This result is obtained using the closed-source model GPT-4.

By contrast, we focus on a comprehensive and unfiltered evaluation setting, considering all theorems contained in the Important Theories, which are explicitly designated by the official seL4 repository and span all seL4 proof categories~\cite{noauthor_sel4_2026,sel4_capdl_2026,sel4_abstract_nodate,sel4_assembly_nodate}. Under this setting, AutoReal-Prover achieves an overall proof success rate of 51.67\%. Notably, AutoReal-Prover is a compact 7B model that supports lightweight local deployment, as required in real-world verification environments, yet still attains substantially higher proof success on seL4.

We further examine the impact of reducing the maximum number of proof attempts per theorem under the same seL4 evaluation setting. With 5 attempts, the success rate is 37.12\%, increasing to 46.36\% with 64 attempts and 50.00\% with 128 attempts, compared to 51.67\% with 256 attempts. This indicates that performance saturates well before the maximum budget, with only marginal gains beyond 64 attempts. In practice, we use 256 attempts to probe the upper bound of LLM-driven proving on difficult real-world theorems, rather than relying on excessive sampling to obtain higher success rates.

\subsubsection{Results on AFP (RQ2)}

Table~\ref{tab:afp-results} reports the proof success rates of AutoReal-Prover on three security-related projects~\cite{RSAPSS-AFP,CRYSTALS-Kyber-Security-AFP,Elliptic-Curves-Group-Law-AFP} from the AFP \cite{noauthor_archive_2026}.
Across these projects, AutoReal-Prover achieves proof success rates ranging from 43.12\% to 59.64\%, with an overall success rate of 53.88\%.
This overall success rate is not lower than that obtained on seL4, indicating that the proving capability of AutoReal-Prover generalizes beyond seL4 to other real-world verification projects.

\begin{table}
\caption{Proof success rates of AutoReal-Prover on AFP security-related projects.}
\label{tab:afp-results}
\centering
\small
\begin{tabularx}{0.92\linewidth}{@{} >{\RaggedRight\arraybackslash}X >{\raggedleft\arraybackslash}p{1.8cm} >{\raggedleft\arraybackslash}p{3cm} @{}}
\toprule
\textbf{AFP project} & \textbf{Solved/Total} & \makecell[r]{\textbf{Proof success}\\\textbf{rate}} \\
\midrule
CRYSTALS-Kyber\_Security            & 32/64    & 50.00\% \\
RSAPSS             & 164/278  & 59.64\% \\
Elliptic\_Curves\_Group\_Law   & 47/109   & 43.12\% \\
\midrule
\textbf{Overall} & \textbf{243/451} & \textbf{53.88\%} \\
\bottomrule
\end{tabularx}
\end{table}

Consistent with the observations on seL4, we further examine how the success rate varies with the number of proof attempts on the AFP projects. The overall success rate reaches 36.36\% with 5 attempts, 51.00\% with 64 attempts, and 52.55\% with 128 attempts, compared to 53.88\% with 256 attempts, suggesting that the reported performance is robust across a range of attempt budgets.

\subsubsection{Quality of Generated Natural-Language Rationales (RQ3)}

We evaluate the quality of the natural-language rationales produced by AutoReal-Prover through a step-level analysis. Specifically, we examine all 670 proof steps and their aligned rationales from the 341 seL4 theorems successfully proved by AutoReal-Prover, with a focus on identifying reasoning errors or mismatches between the rationales and the corresponding proof commands. We find that inaccurate rationales are concentrated in a small number of structurally specific cases, most notably the terminal \texttt{done} step. Among these, 84 rationales associated with \texttt{done} contain irrelevant or incorrect explanations, often involving unrelated lemmas or theories. This is likely due to the high frequency and syntactic uniformity of \texttt{done} in seL4 proofs, which makes this semantically underspecified step prone to explanation hallucination. Outside of \texttt{done}, only 10 additional inaccurate rationales are observed, mainly involving minor mismatches for \texttt{by simp}, overly generic explanations for \texttt{apply auto}, and spurious references under \texttt{apply wpsimp+}. Excluding \texttt{done}, the inaccuracy rate is 1.49\%.


Beyond correctness, we qualitatively examine the usefulness of the generated rationales for human verifiers in Appendix~\ref{appendix:proof_rationales}. The case study compares rationales produced for correct and incorrect proofs of the same theorem, and shows that they exhibit different structural characteristics, which can assist verifiers in identifying deviations in failed proof attempts and informing subsequent repair.

Overall, although the generated rationales are not perfect and exhibit occasional systematic errors in specific structural cases, they are largely accurate and informative, providing support for proof inspection and repair.

\subsubsection{Comparison with Human-Written Proofs (RQ4)}

\begin{figure}[H]
\begin{center}
	\includegraphics[width=0.65\textwidth]{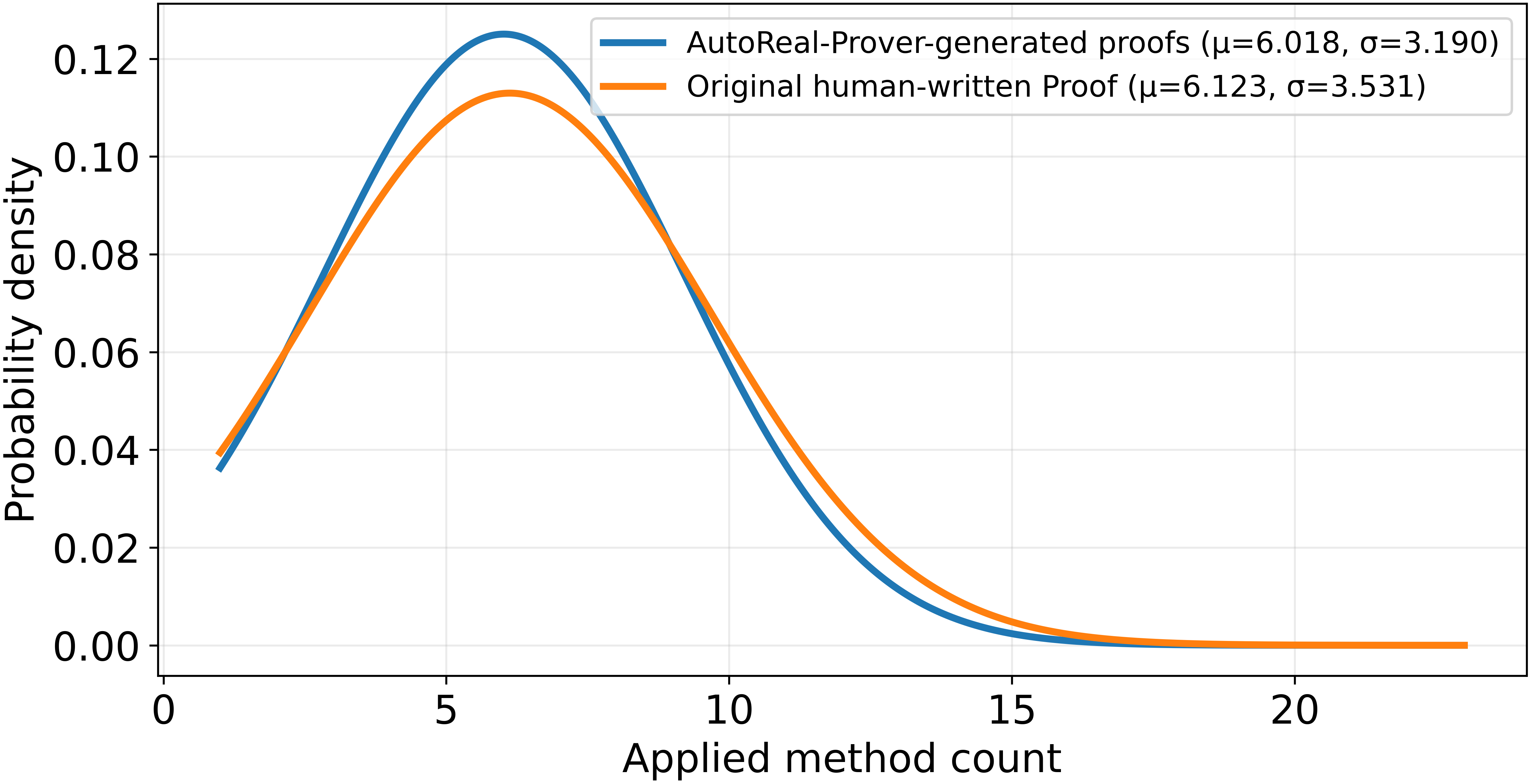}
	\caption{Distribution of applied method counts.}
	\label{fig:method_count_dist}
\end{center}
\end{figure}

We compare proofs generated by AutoReal-Prover with the original human-written proofs for the 341 theorems successfully proved in the seL4 evaluation, along two dimensions: proof duplication and method usage. Among these theorems, 93.55\% of the generated proofs differ from their human-written counterparts, indicating that the model does not reproduce existing proofs.
We further compare the proofs by counting the number of applied methods per proof, including Isabelle proof methods as well as referenced lemmas and definitions. AutoReal-Prover-generated proofs use 6.018 methods on average, compared to 6.123 methods in the original proofs. Figure~\ref{fig:method_count_dist} shows the corresponding distributions. Appendix~\ref{appendix:generated_vs_human} analyzes differences in proof methods using a single example.

These results indicate that most AutoReal-Prover-generated proofs differ from the original human-written proofs, demonstrating their distinctness.

\subsubsection{Ablation Study (RQ5)}
To assess whether the two key improvements introduced in this work are responsible for the gains in proof success rates, we conduct an ablation study on (1) CoT-based proof training and (2) context augmentation. We use exactly the same evaluation configuration as in prior experiments, and evaluate on a subset of the seL4 test set: the 105 theorems from the \texttt{invariant-abstract} category (\texttt{Syscall\_AI.thy})~\cite{sel4_abstract_nodate}. Table~\ref{tab:ablation-syscall-ai} summarizes the results.

\begin{table}
\caption{Ablation study.}

\label{tab:ablation-syscall-ai}
\centering
\small
\begin{tabularx}{\linewidth}{@{}
  >{\RaggedRight\arraybackslash}X
  >{\raggedleft\arraybackslash}m{2.2cm}
  >{\raggedleft\arraybackslash}m{3cm}
@{}}
\toprule
\textbf{Method} & \textbf{Solved/Total} & \makecell[r]{\textbf{Proof success}\\\textbf{rate}} \\
\midrule
Base model (Qwen2.5-Coder-7B) & 1/105   & 0.95\%  \\
Base model + CoT-based proof training & 24/105  & 22.86\% \\
Base model + context augmentation & 6/105   & 5.71\%  \\
\midrule
\parbox[c]{0.95\linewidth}{\textbf{Our method}\\{\footnotesize (Base model + CoT-based proof training + context augmentation)}}
& \textbf{57/105} & \textbf{54.29\%} \\
\bottomrule
\end{tabularx}
\end{table}

As shown in Table~\ref{tab:ablation-syscall-ai}, the base model solves only 1 out of 105 theorems. Adding either CoT-based proof training or context augmentation increases the proof success rate, but the performance remains limited when applying only one component. When both improvements are enabled, the success rate increases substantially, reaching 54.29\% (57/105), indicating that the two components are complementary and together account for the observed gains.



\section{Related Work}\label{sec:related}
Formal verification provides rigorous techniques for establishing system correctness, with mature paradigms such as model checking and theorem proving~\cite{clarke1996formal,nawaz2019survey,Wang2019FormalMethods,clarke1997model,nipkow2014concrete}. Within theorem proving, prior work has pursued different directions for improving automation. Fully automated, SMT-style approaches are effective in restricted domains but limited in expressiveness for rich program semantics~\cite{sutcliffe2001evaluating,de2008z3}. In contrast, interactive theorem proving combines automated reasoning with human guidance to support verification in expressive logical frameworks~\cite{viswanathan2024automating,nipkow2002isabelle,moura2021lean}. This paper focuses on automating interactive theorem proving, and reviews prior work in this direction, with emphasis on proof automation and recent LLM-based techniques for reducing human effort in large-scale verification.

\paragraph{Interactive Theorem Proving~(ITP).}
Proof assistants such as Isabelle~\cite{nipkow2002isabelle}, Coq~\cite{huet1997coq}, and Lean~\cite{moura2021lean} support machine-checked proofs through a combination of human guidance and automated reasoning. To reduce manual effort, these systems provide various automation mechanisms. In Isabelle, Sledgehammer integrates external automated provers into the interactive proving workflow~\cite{blanchette2011automatic,blanchette2013extending}, while further work has improved automation through algebraic methods~\cite{guttmann2011automating}. Similar automation efforts have also been explored in other proof assistants, such as SMT-based automation in Coq~\cite{ekici2017smtcoq}.  
Despite these advances, existing automation remains insufficient for large-scale, real-world verification. In industrial projects such as seL4~\cite{heiser2020sel4}, proof development continues to dominate verification cost, with human effort as the primary scalability bottleneck~\cite{klein2009sel4,gu2016certikos}.

\paragraph{LLM-Driven Theorem Proving.}
Large language models (LLMs) have demonstrated strong reasoning capabilities across domains, motivating their application to automated theorem proving~\cite{lama2024benchmarking}. Early work by Polu et al.~\cite{polu2020generative} introduced GPT-f for Metamath, showing that LLMs can generate formal proofs. Zheng et al.~\cite{zheng2021minif2f} later proposed miniF2F, a multilingual benchmark supporting proof assistants such as Isabelle and Lean, which has since become the standard testbed for evaluating LLM-driven theorem proving.
Most subsequent work focuses on improving proof success rates on miniF2F-test. In the Isabelle track, representative systems include DSP~\cite{jiang2022draft} (39.3\%), LEGO-Prover~\cite{wang2023lego} (50.02\%), HybridProver~\cite{hu2025hybridprover} (59.4\%), and ProofAug~\cite{liu2025proofaug} (66.0\%). In the Lean track, substantially higher success rates have been reported, for example DeepSeek-Prover-V2~\cite{ren2025deepseek} (88.9\%), Delta Prover~\cite{zhou2025solving} (95.9\%), and Seed-Prover~\cite{chen2025seed} (99.6\%).
While these approaches advance the state of the art on benchmark datasets that primarily target mathematical problem solving, real-world verification poses different challenges. Proofs in industrial verification rely on extensive project-specific context and focus on complex system security properties~\cite{klein2009sel4,klein2014comprehensive,zhang2024selene,lin2024fvel}. Bridging this gap remains a challenge for applying LLM-driven theorem proving to real-world verification.

\paragraph{LLM-Driven Theorem Proving for Real-World Verification.}
Recent work has begun to explore LLM-driven theorem proving in real-world verification settings~\cite{lin2024fvel,zhang2024selene,qin2025can,bayazit2025case}. Among studies targeting seL4, Lin et al.~\cite{lin2024fvel} constructed the FVELER dataset by extracting theorems, proof steps, and proof states from seL4, and evaluated automated proof attempts on this dataset, reporting proof success rates between 5.75\% and 8.17\% on the FVELER test and test-hard splits. Zhang et al.~\cite{zhang2024selene} presented Selene as an early attempt at automated proof generation for seL4, evaluating 340 sampled theorems across multiple sessions and reporting a best averaged proof success rate of 27.06\%. These results highlight both the potential of LLM-driven theorem proving for real-world verification and the remaining performance gap. Moreover, existing approaches to industrial-scale verification often rely on closed-source, extremely large models that cannot be locally deployed~\cite{zhang2024selene,qin2025can,bayazit2025case,curaba2cryptoformaleval}, while practical settings demand lightweight and deployable solutions~\cite{paloniemi2025porting,pan2025cost,ye2024enabling}. 
Building on this line of work, our study targets higher proof success rates while supporting lightweight local deployment.

\section{Conclusion and Future Work}\label{sec:conclusion}

In this paper, we presented AutoReal, an LLM-driven approach to automated theorem proving for real-world industrial-scale verification, and evaluated it on the seL4-Isabelle verification project. AutoReal integrates proof chain-of-thought training with targeted context augmentation to support automated proof construction while making proof reasoning explicit. Based on this approach, we fine-tuned Qwen2.5-Coder-7B to obtain AutoReal-Prover. Under a comprehensive and unfiltered evaluation on seL4, AutoReal-Prover achieves a 51.67\% proof success rate, substantially improving over prior results, and evaluation on security-related AFP developments indicates good generality. AutoReal-Prover produces step-aligned explanations alongside generated proofs to support proof understanding. All results are obtained using a compact 7B-scale model, enabling cost-efficient local deployment under practical industrial constraints.

Overall, this work advances the practical applicability of LLM-driven theorem proving in real-world industrial-scale verification.


\paragraph{Future work.}
Although AutoReal is implemented and evaluated in Isabelle/HOL, the approach is not Isabelle-specific. It is grounded in real-world verification workflows, where proofs rely on large and fragmented proof contexts accumulated during development.
This characteristic is shared by large-scale verification projects in other interactive theorem provers such as Coq and Lean, including systems like CompCert~\cite{leroy2016compcert}.
A natural direction for future work is therefore to extend the approach to other real-world verification projects, including those developed in different formal languages. At the same time, while AutoReal closely follows existing verification practice, the extraction of theorem-relevant proof context is not yet automated and currently requires human input. Future work will focus on automating proof-context extraction and improving proof success rates on industrial-scale verification tasks, with the aim of further reducing manual effort in real-world formal proof development.

\bibliographystyle{splncs04}
\bibliography{7.mybibliography}

\appendix
\section*{Appendix}

\section{Prompt Templates}
\label{appendix:prompt-templates}

\subsection{Prompt for Proof Generation}
\phantomsection
\label{appendix:prompt_proof_generation}
The prompt shown in Fig.~\ref{fig:prompt-proof} is used to generate a complete Isabelle proof script together with step-aligned rationales.

\begin{figure}[H]
\centering
\begin{minipage}{1\textwidth}
\raggedright
\begin{minted}[
  frame=none,
  fontsize=\fontsize{8}{9}\selectfont,
  breaklines,
  tabsize=2
]{text}
You are an Isabelle proof expert, and I need you to help me prove the theorems in the project and explain the reasoning behind each proof step. I will provide the lemma to be proved and relevant proof context. Based on these details, you should construct the complete Isabelle proof step by step, and after each proof step you must write out the reasoning (chain of thought) for that step. Your final answer should contain the full sequence of proof steps together with their corresponding chain-of-thought explanations.
\end{minted}
\end{minipage}
\caption{Prompt for proof generation.}
\label{fig:prompt-proof}
\end{figure}

\subsection{Prompt for CoT Construction}
\phantomsection
\label{appendix:prompt_cot_construction}
The prompt shown in Fig.~\ref{fig:prompt-cot} is used in our proof chain-of-thought construction stage in Section~\ref{CoT_data_construction} to generate CoT instances.

\begin{figure}[H]
\centering
\begin{minipage}{1\textwidth}
\raggedright
\begin{minted}[
  frame=none,
  fontsize=\fontsize{8}{9}\selectfont,
  breaklines,
  tabsize=2
]{text}
You are analyzing a single proof step from the seL4 Isabelle verification project. Please combine the following information and explain the reasoning behind this proof step and its role in the larger proof.
TASK:
1. Read the following `proof_step`(the content of the proof step), `proof_state_before_applying_step`, `proof_state_after_applying_step`, and `source_lemma`(which lemma this proof step comes from).
2. Explain the reasoning of THIS SINGLE proof step.
HARD CONSTRAINTS:
- Do NOT simply restate or paraphrase the input text; focus on the reasoning, methods used, and how the proof state changes.
- Write ONE single concise paragraph in a technical, third-person style. Do NOT use phrases like "I think", "I am trying", or "let me explain".
\end{minted}
\end{minipage}
\caption{Prompt for CoT construction.}
\label{fig:prompt-cot}
\end{figure}

\section{Examples}
\phantomsection
\label{appendix:examples}
This section presents representative artifacts generated in this study, together with supporting analyses.

\subsection{Example CoT Instance}
\phantomsection
\label{appendix:cot_instance}
Figure~\ref{fig:cot_instance} shows a complete CoT instance for a single proof step, produced by our proof CoT data construction procedure in Section~\ref{CoT_data_construction}.

\begin{figure}[H]
\begin{center}
\begin{minipage}{1\textwidth}

\begin{minted}[
  frame=none,
  encoding=utf8,
  fontsize=\fontsize{8}{9}\selectfont,
  breaklines,
  tabsize=2
 ]{isabelle}
(** Source lemma **)
lemma power_2_mult_step_le:
  "⟦n' ≤ n; 2 ^ n' * k' < 2 ^ n * k⟧ ⟹ 2 ^ n' * (k' + 1) ≤ 2 ^ n * (k::nat)"
(** Proof state before applying this step **)
proof (prove)
goal (1 subgoal):
 1. ⟦n' ≤ n; 2 ^ n' * k' < 2 ^ n * k⟧ ⟹ 2 ^ n' * (k' + 1) ≤ 2 ^ n * k
(** Proof step **)
apply (cases "n' = n", simp)
(** Proof state after applying this step **)
proof (prove)
goal (2 subgoals):
 1. ⟦k' < k; n' = n⟧ ⟹ 2 ^ n + 2 ^ n * k' ≤ 2 ^ n * k
 2. ⟦n' ≤ n; 2 ^ n' * k' < 2 ^ n * k; n' ≠ n⟧ ⟹ 2 ^ n' * (k' + 1) ≤ 2 ^ n * k
(** CoT Rationale **)
(*The proof step applies the `cases` tactic to split the goal into two cases based on whether `n'` equals `n`. The first case assumes `n' = n`, simplifying the inequality directly. The second case assumes `n' < n`, leading to a new subgoal that must be proven separately. This approach allows the proof to handle both scenarios where `n'` is equal to or less than `n`, ensuring comprehensive coverage of the original goal.*)

\end{minted}
\end{minipage}
\end{center}
\caption{A CoT instance for a single proof step.}
\label{fig:cot_instance}
\end{figure}

\subsection{Example: Step-Aligned Rationales for a Successful and a Failed Proof}
\phantomsection
\label{appendix:proof_rationales}

Beyond proof correctness, this appendix provides a qualitative case study of how step-aligned rationales can support human verification.
We compare a successful proof and a failed proof attempt generated by AutoReal-Prover for the same lemma \texttt{sameFor\_reads\_equiv\_f\_g} in \texttt{Noninterference.thy}.
In both cases, each Isabelle command is paired with a step-level rationale that explains which definitions or lemmas are being exploited, how the proof state is expected to evolve, and why the tactic choice is locally appropriate.
Figures~\ref{fig:successful_proof_rationales} and~\ref{fig:failed_proof_rationales} present the two scripts together with their rationales.

In the successful case (Fig.~\ref{fig:successful_proof_rationales}), the rationales remain aligned with the proof’s progression.
They begin by describing a controlled normalization step that unfolds the key definitions \texttt{reads\_equiv\_f\_g\_def}, \texttt{sameFor\_def}, and \texttt{reads\_equiv\_def2}, and they emphasize the role of \texttt{if\_cong} in simplifying conditional structure without triggering premature branching.
They then motivate a cleanup step using \texttt{clarsimp}, while explicitly disabling \texttt{if\_split} to avoid case explosion after unfolding.
Finally, they justify the closing automation with \texttt{fastforce} and point to the supporting facts \texttt{silc\_dom\_equiv\_def} and \texttt{pasSubject\_not\_SilcLabel} that discharge the remaining obligations.
Overall, the successful rationales form a clear flow from normalization, to simplification, to automated closure.

In contrast, the failed attempt (Fig.~\ref{fig:failed_proof_rationales}) exhibits structural mismatches that become visible from the rationales.
It starts with an aggressive \texttt{clarsimp} that unfolds multiple definitions and domain facts at once, which can leave a more complex conditional and quantified goal structure for later steps.
The subsequent \texttt{arg\_cong} step is framed as a transformation through \texttt{internal\_state\_if}, yet its rationale does not establish a clear connection to the remaining obligations about membership in \texttt{same\_for} and the partition induced by \texttt{label\_of (pasSubject aag)}.
The final \texttt{auto} step is justified only at a high level, and the script terminates with \texttt{done} while subgoals remain.
Taken together, the rationales suggest that the failure arises from an early divergence followed by insufficiently targeted follow-up steps, rather than from a single missing lemma at the end.

\paragraph{Implications for inspection and repair.}
Comparing the two cases helps a verifier localize where the failed attempt departs from a viable strategy, namely the initial aggressive simplification and the subsequent unmotivated detour.
The contrast also suggests concrete repairs: reintroduce congruence-guided normalization by including \texttt{cong: if\_cong}, control conditional splitting with \texttt{split del: if\_split}, and ensure that the final automation is supported by the same domain and label facts that enable closure in the successful script.
Together, this example illustrates that correct and incorrect rationales can exhibit distinct structural patterns, and that these patterns help verifiers identify deviations in failed attempts and guide subsequent repair.

\begin{figure}[H]
\begin{center}
\begin{minipage}[t]{1\textwidth}
\centering
\textsc{Successful Whole Proof with Step-Level Rationales}\\[-0.6em]
\rule{0.9\linewidth}{0.2pt}
\begin{minted}[
  frame=none,
  fontsize=\scriptsize,
  breaklines,
  tabsize=2
]{isabelle}
lemma sameFor_reads_equiv_f_g:
  "⟦ reads_equiv_f_g aag s s'; silc_inv aag st' st'';
     pasSubject aag ∈ pasDomainAbs aag (cur_domain s) ∪ pasDomainAbs aag (cur_domain s') ⟧
   ⟹ (((uc,s),mode),((uc,s'),mode)) ∈ same_for aag (Partition (label_of (pasSubject aag)))"
  apply (simp add: reads_equiv_f_g_def sameFor_def sameFor_subject_def reads_equiv_def2 cong: if_cong)
  (* The proof step applies simplification rules (`simp`) to the goal using definitions and congruence lemmas related to `reads_equiv_f_g`, `sameFor`, `sameFor_subject`, and `reads_equiv` (in particular `reads_equiv_def2`), together with the congruence rule `if_cong`. Intuitively, the statement connects a read-equivalence assumption (`reads_equiv_f_g`) with a membership claim in `same_for` over a partition determined by the subject’s label. Unfolding `reads_equiv_f_g_def` and the `sameFor_*` definitions exposes the underlying logical structure of “same-for” in terms of observations that matter to the subject/partition. Using `cong: if_cong` is crucial because these definitions typically contain conditional branches (e.g., depending on whether an object is in the subject’s domain/label or on mode/domain conditions). The congruence allows `simp` to simplify both sides of `if` expressions in a goal-directed way without losing equivalence. Overall, this step normalizes the goal by expanding the relevant notions and rewriting them into a simpler, more explicit form, reducing the problem to a smaller set of concrete logical obligations about the two states `s` and `s'` under the given domain assumption. *)
  apply (clarsimp split del: if_split)
  (* The proof step applies `clarsimp` to aggressively simplify and clean up the proof state: it combines classical reasoning (`clarify`) with rewriting/simplification and will also introduce the most useful quantified assumptions and split conjunctions. The modifier `split del: if_split` explicitly disables the default splitting of `if` expressions. This is a deliberate choice: after the previous unfolding, there can be many nested `if`s, and blindly splitting them can explode the number of cases. By preventing `if_split`, `clarsimp` focuses on simplifying the goal using the available assumptions instead of branching prematurely. As a result, this step typically turns the expanded “same-for / reads-equivalence” obligations into a small number of clean subgoals (often universally quantified or implication-shaped), where irrelevant syntactic noise is removed and the remaining conditions are expressed in a form suitable for automated solvers. *)
  by (fastforce simp: silc_dom_equiv_def pasSubject_not_SilcLabel)+
  (* The final step discharges the remaining subgoals using `fastforce`, an automated tactic that combines rewriting, classical reasoning, and a search procedure to solve goals efficiently. The `simp:` context supplies two key ingredients: - `silc_dom_equiv_def`: expands/simplifies the notion of SILC domain equivalence that follows from the `silc_inv` assumption, making domain-related constraints explicit and usable by automation. -`pasSubject_not_SilcLabel`: provides a domain/label side-condition excluding the subject from the SILC label (or an analogous “not-in” fact), which helps rule out problematic cases in the `same_for`/partition reasoning. With these simplifications available, `fastforce` can close the goals by resolving the remaining logical dependencies between read-equivalence, domain membership, and the partition induced by `label_of (pasSubject aag)`. The trailing `+` repeats the method as needed for any remaining subgoals generated earlier, ensuring all goals are solved. *)
\end{minted}
\end{minipage}
\end{center}
\caption{A successfully verified proof with step-aligned rationales.}
\label{fig:successful_proof_rationales}
\end{figure}

\begin{figure}[H]
\begin{center}
\begin{minipage}[t]{1\textwidth}
\centering
\textsc{Failed Proof Attempt with Step-Level Rationales}\\[-0.6em]
\rule{0.9\linewidth}{0.2pt}
\begin{minted}[
  frame=none,
  fontsize=\scriptsize,
  breaklines,
  tabsize=2
]{isabelle}
lemma sameFor_reads_equiv_f_g:
  "⟦ reads_equiv_f_g aag s s'; silc_inv aag st' st'';
     pasSubject aag ∈ pasDomainAbs aag (cur_domain s) ∪ pasDomainAbs aag (cur_domain s') ⟧
   ⟹ (((uc,s),mode),((uc,s'),mode)) ∈ same_for aag (Partition (label_of (pasSubject aag)))"
apply (clarsimp simp: reads_equiv_def2 sameFor_def sameFor_subject_def
                        silc_dom_equiv_def pasSubject_not_SilcLabel)
(* This attempt starts with an aggressive `clarsimp` that expands `reads_equiv_def2` and the `sameFor_*` definitions, while also unfolding domain-related facts via `silc_dom_equiv_def` and `pasSubject_not_SilcLabel`. Compared with a more controlled normalization (e.g., first using `simp ... cong: if_cong` and postponing splits), this single step can over-simplify or expose nested conditionals/quantifiers in a form that is harder for later automation to finish. In particular, unfolding without guiding congruence rules may leave conditional branches in a less canonical shape, and the resulting subgoals may retain subtle dependencies that `auto` does not resolve. *)
apply (drule_tac f="internal_state_if" in arg_cong[where P=id])
(* This step tries to rewrite an equality by applying `arg_cong` to both sides under the function `internal_state_if`. However, in this proof, the key remaining obligations are about membership in `same_for` for a specific partition and the relationship induced by `reads_equiv_f_g`. Lifting the goal through `internal_state_if` is not obviously aligned with those obligations, so it may transform the goal into a form that does not match the available lemmas, or it may introduce an unnecessary detour that blocks straightforward closure. *)
apply (auto intro!: states_equiv_for_sym[OF reads_equiv_f_g] elim!: bexI[rotated] bexE[rotated] allE simp: fun_upd_idem)[1]
(* The attempt then uses `auto` with selected intro/elim rules and simplifications. While these rules are broadly useful, they are not sufficiently targeted to discharge the remaining subgoals produced by the earlier `clarsimp`/`arg_cong` steps. Typically, this stage still leaves residual obligations about the exact “same-for” observation constraints for the subject’s partition (often involving conditionals or domain cases). As a result, Isabelle will report remaining subgoals rather than completing the proof. *)
done
(* The script ends with `done`, but this attempt is marked as failed: the preceding automation does not fully discharge all subgoals, so Isabelle would not accept the proof at this point. *)
\end{minted}
\end{minipage}
\end{center}
\caption{A failing proof attempt with step-aligned rationales.}
\label{fig:failed_proof_rationales}
\end{figure}

\subsection{Example Comparison Between Generated Proof and Original Proof}
\phantomsection
\label{appendix:generated_vs_human}

Figure~\ref{fig:generated_vs_human} presents a comparison between an AutoReal-Prover-generated proof and the original human-written proof for the same theorem. While both scripts discharge the same goal, they differ in structure and tactic composition. In addition, the generated script includes step-aligned natural-language rationales (shown as comments) that describe the intended role of each step.

In the original proof, the human-written script normalizes the goal early: \texttt{clarsimp} with \texttt{has\_reply\_cap\_def} expands the predicate and simplifies the Hoare triple into preservation obligations expressed in terms of the concrete conditions that characterize reply capabilities. The remaining work is then handled by \texttt{wpsimp} over \texttt{thread\_set}, where the \texttt{thread\_set}-specific lemma \texttt{thread\_set\_cte\_wp\_at\_trivial} and the WP/VCG infrastructure (including \texttt{hoare\_vcg\_all\_lift}) provide the key preservation facts needed for automation to finish.

In contrast, the generated proof separates these steps more explicitly. It unfolds \texttt{has\_reply\_cap\_def} first, applies \texttt{hoare\_pre} as an explicit precondition-adjustment step before automation, and then closes the goal with \texttt{wpsimp} by relying on more general Hoare/WP infrastructure, such as \texttt{hoare\_drop\_imps}, \texttt{cte\_wp\_at\_caps\_of\_state}, and \texttt{hoare\_vcg\_all\_lift}, to discharge the resulting side conditions.

In this example, the two scripts use different step decompositions and tactic combinations to reach the same verified goal, suggesting a modest but concrete divergence in proof ideas.

\begin{figure}[H]
\begin{center}

\begin{minipage}[t]{0.46\textwidth}
\centering
\textsc{Original Proof}\\[-0.6em]
\rule{0.9\linewidth}{0.2pt}
\begin{minted}[frame=none,fontsize=\scriptsize,breaklines,tabsize=2]{isabelle}
lemma thread_set_has_no_reply_cap:
  "(⋀tcb. ∀(getF, v)∈ran tcb_cap_cases. getF (f tcb) = getF tcb)
   ⇒ ⦃λs. ¬has_reply_cap tt s⦄ thread_set f t ⦃ λ_ s. ¬has_reply_cap tt s⦄"
  apply (clarsimp simp add: has_reply_cap_def)
  apply (wpsimp wp: hoare_vcg_all_lift thread_set_cte_wp_at_trivial | fast)+
  done
\end{minted}
\end{minipage}
\hfill
\vrule{}
\hfill
\begin{minipage}[t]{0.52\textwidth}
\centering 
\textsc{AutoReal-Prover-Generated Proof}\\[-0.6em]
\rule{0.9\linewidth}{0.2pt}
\begin{minted}[frame=none,fontsize=\scriptsize,breaklines,tabsize=2]{isabelle}
lemma thread_set_has_no_reply_cap:
  "(⋀tcb. ∀(getF, v)∈ran tcb_cap_cases. getF (f tcb) = getF tcb)
   ⇒ ⦃λs. ¬has_reply_cap tt s⦄ thread_set f t ⦃ λ_ s. ¬has_reply_cap tt s⦄"
  unfolding has_reply_cap_def
  (* The proof step unfolds the definition of `has_reply_cap` using the provided lemma... *)
  apply (rule hoare_pre)
  (* The proof step applies the `hoare_pre` rule to the goal, transforming the precondition into... *)
  apply (wpsimp wp: hoare_drop_imps simp: cte_wp_at_caps_of_state hoare_vcg_all_lift)+
  (* The proof step applies simplification rules (`wpsimp`) and specific lemmas `hoare_drop_imps... *)
  done
  (* The proof step completes successfully, indicating that the goal has been proven... *)
\end{minted}
\end{minipage}
\end{center}
\caption{Comparison Between Generated Proof and Original Proof.}
\label{fig:generated_vs_human}
\end{figure}

\section{Training Configuration}
\phantomsection
\label{appendix:training_details}

This appendix details the training configurations corresponding to the CoT-based proof training procedure described in Section~\ref{proof_training}.

Training is performed in bf16 precision with a maximum sequence length of 8192.
We use a per-device batch size of 1 with gradient accumulation of 32, resulting in an effective batch size of 32, and train the model for 10 epochs.
Optimization uses a cosine learning-rate schedule with a warmup ratio of 0.03, a learning rate of $5\times10^{-6}$, weight decay 0.05, a maximum gradient norm of 0.3, and the paged AdamW 8-bit optimizer.

\section{Evaluation Theorem Sets}
\phantomsection
\label{appendix:evaluation-theorem-sets}

\subsection{seL4 Proof Categories and Important Theories}
\phantomsection
\label{appendix:sel4-proof-categories}

The seL4 microkernel verification is a large-scale mechanized proof effort in Isabelle/HOL.
The seL4 verification repository organizes its proof development into ten logical modules that mirror the repository directory structure.
Across these modules, the seL4 verification connects high-level specifications to lower-level models and implementations, and it establishes key system properties such as invariant preservation, access control, CapDL-based initialization reasoning, and information-flow security.

\begin{enumerate}
  \item \textbf{access-control.}
  This category proves access control properties at the abstract level.
  Building on global invariants, it shows that authority is mediated by capabilities and that state changes respect permission constraints.
  An important entry theory is \texttt{Syscall\_AC.thy}, which develops syscall-level reasoning used to establish preservation of access control invariants.

  \item \textbf{asmrefine.}
  This category supports the binary-oriented part of the verification chain by relating C-level execution to a graph-level representation used in subsequent low-level reasoning.
  It establishes refinement-style correspondence between the C semantics and the exported control-flow graph semantics.
  An important entry theory is \texttt{SEL4GraphRefine.thy}, which contains key refinement statements for this correspondence.

  \item \textbf{bisim.}
  This category develops bisimulation-style reasoning to justify strong separation/isolation properties for restricted, static configurations.
  The results support a separation-kernel interpretation where permitted interactions are tightly controlled by policy and kernel mechanisms.
  An important entry theory is \texttt{Separation.thy}, which formalizes core definitions and theorems used in this isolation argument.

  \item \textbf{capDL-api.}
  This category connects kernel API behavior to CapDL-level initialization reasoning and supports system bring-up proofs with modular heap-style arguments.
  It provides lemmas that track how kernel operations affect abstract heap structures and initialization-relevant state.
  An important entry theory is \texttt{KHeap\_DP.thy}, which develops preservation lemmas for heap-related invariants used in this setting.

  \item \textbf{crefine.}
  This category proves that the C implementation refines the executable design-level model.
  Together with the higher-level refinement links, it establishes that the kernel’s C behavior is consistent with the intended specification.
  An important entry theory is \texttt{Refine\_C.thy}, which contains the central refinement results connecting the design model to the C kernel.

  \item \textbf{drefine.}
  This category establishes design-level refinement results that relate the abstract specification to intermediate, executable models used in the verification stack.
  These links provide the bridge that enables properties proved at higher levels to be transferred to more concrete models.
  An important entry theory is \texttt{Refine\_D.thy}, which develops the main correspondence results for this layer.

  \item \textbf{infoflow.}
  This category establishes information-flow security results at the abstract level and transfers them through refinement to more concrete levels where applicable.
  The proofs capture noninterference-style confidentiality guarantees under an explicit policy.
  An important entry theory is \texttt{Noninterference.thy}, which develops the core information-flow security theorems.

  \item \textbf{invariant-abstract.}
  This category proves global invariants of the abstract specification using monadic Hoare logic.
  The results show that kernel entry points preserve key invariants across system calls and fault handling.
  An important entry theory is \texttt{Syscall\_AI.thy}, which contains central invariant-preservation lemmas used throughout this category.

  \item \textbf{refine.}
  This category establishes refinement between the abstract specification and the executable design-level model while maintaining design-level invariants needed for later stages.
  It provides the main refinement infrastructure that underpins transfer of correctness properties down the stack.
  An important entry theory is \texttt{Refine.thy}, which contains core refinement theorems for this layer.

  \item \textbf{sep-capDL.}
  This category builds a separation-logic style framework for the CapDL specification, enabling modular reasoning about disjoint resources and their composition.
  The development supports initialization and configuration proofs by structuring state into separable components.
  An important entry theory is \texttt{AbstractSeparation\_SD.thy}, which instantiates the underlying separation-algebraic reasoning principles used in this category.
\end{enumerate}

\subsection{AFP Projects}
\phantomsection
\label{appendix:afp-projects}

The Archive of Formal Proofs (AFP) is a community-maintained repository of Isabelle/HOL developments across mathematics and computer science.
To evaluate AutoReal beyond operating-system verification, we select three AFP developments that stress different proof characteristics, including standard-driven cryptographic reasoning and algebra-heavy mathematical formalization.

\begin{enumerate}
  \item \textbf{RSAPSS.}
  This project formalizes the RSA-PSS signature scheme in Isabelle/HOL and develops the associated correctness and security-oriented reasoning in the style of standard cryptographic constructions.
  It provides a representative target for proofs that combine specification structure with probabilistic or reduction-style arguments.

  \item \textbf{Elliptic\_Curves\_Group\_Law.}
  This project formalizes the group law for Weierstrass elliptic curves and extends the reasoning to efficient coordinate representations over prime fields.
  It is dominated by algebraic structure, non-trivial case distinctions, and layered abstractions typical of mechanized mathematics in cryptography.

  \item \textbf{CRYSTALS-Kyber\_Security.}
  This project formalizes security reasoning for CRYSTALS\-Kyber within the CryptHOL-based probabilistic framework, including probabilistic correctness and standard indistinguishability-style security notions.
  It represents modern post-quantum cryptographic verification, where randomized algorithms and game-based arguments play a central role.
\end{enumerate}



\end{document}